# CMD: A Multi-Channel Coordination Scheme for Emergency Message Dissemination in IEEE 1609.4

Odongo Steven Eyobu[1, 2], Jhihoon Joo[1], Dong Seog Han[1]

[1] School of Electronics Engineering, Kyungpook National University, 80 Daehak-ro, Buk-gu, Daegu, 41566, Republic of Korea
[2] School of Computing & Informatics Technology, Makerere University, Plot 56, Pool Road, P.O. Box 7062, Kampala, Uganda

Correspondence should be addressed to Dong Seog Han; dshan@knu.ac.kr

**Abstract**
In the IEEE 1609.4 legacy standard for multi-channel communications in vehicular ad hoc networks (VANETs), the control channel (CCH) is dedicated to broadcast safety messages while the service channels (SCHs) are dedicated to transmit infotainment service content. However, the SCH can be used as an alternative to transmit high priority safety messages in the event that they are invoked during the service channel interval (SCHI). This implies that there is a need to transmit safety messages across multiple available utilized channels to ensure that all vehicles receive the safety message. Transmission across multiple SCH's using the legacy IEEE 1609.4 requires multiple channel switching and therefore introduces further end-to-end delays. Given that safety messaging is a life critical application, it is important that optimal end-to-end delay performance is derived in multi-channel VANET scenarios to ensure reliable safety message dissemination. To tackle this challenge, three primary contributions are in this article: first, a channel coordinator selection approach based on the least average separation distance (LAD) to the vehicles that expect to tune to other SCH's and operates during the control channel interval (CCHI) is proposed. Second, a model to determine the optimal time intervals in which CMD operates during the CCHI is proposed. Third, a contention back-off mechanism for safety message transmission during the SCHI is proposed. Computer simulations and mathematical analysis show that CMD performs better than the legacy IEEE 1609.4 and a selected state-of-the-art multi-channel message dissemination scheme in terms of end-to-end delay and packet reception ratio.

## 1. Introduction

Nowadays, intelligent transport systems (ITS) are one of the key drivers for the evolution of smart cities. Among the major enabling technologies to realize this evolution is vehicular communications technology (VCT). VCTs should be able to provide services such as safety on the road and in-vehicle on-demand infotainment content. The IEEE 1609.4 standard [1] is the basic technology designed to achieve and enable the implementation of both cooperative safety message dissemination and provision of infotainment services through multi-channel communications. Seven 10 MHz channels have been reserved in the 5.9 GHz frequency band [2] for this purpose.

The multi-channels defined therein are the control channel (CCH) and six service channels (SCHs) all operating at fixed intervals. The CCH is dedicated to broadcast safety messages while the SCHs are dedicated to transmit infotainment service content. During the CCH interval (CCHI), all vehicles must tune to the CCH unlike during the SCH interval (SCHI). Furthermore, the standard defines the continuous and alternating channel access modes. In the continuous channel access mode, vehicles tune to the CCH until they demand for a service that has been advertised. The alternating channel access mode allows vehicles to always switch between the CCH and their desired advertised SCH after an interval of 50 ms.

When the different vehicles switch to their desired SCH's during the SCHI, it limits the possibility of transmitting safety broadcast messages to all vehicles in the event of an emergency during the SCHI. This is a threat to the reliability of safety message transmission especially because further end-to-end delays are introduced. Therefore, it is necessary to design inter-channel communication mechanisms across service channels which should be able to meet requirements such as minimum end-to-end delay for emergency safety message transmission and delivery.

Various studies [3-11] have proposed approaches on improving end-to-end delay performance for vehicular ad-hoc network (VANET) in multi-channel conditions. The major considerations in these previous studies include: (1) using channel coordination vehicles [6] (2) using road side units (RSUs) as coordinators [10, 11] (3) dynamic variable CCHI and SCHI [5] and (4) time slot utilization based on peer to peer negotiation as a multi-channel coordination function [4]. A detailed review of studies [3-11] is covered in section II. However for the purpose of this study, the wireless access to vehicular environments – enhanced safety message delivery approached (WSD) [6] is used for comparison with the proposed scheme. During the CCHI, in the WSD approach, each vehicle collects data including the expected SCH that the vehicles in its communication range expect to tune to during the SCHI, computes the delay in each SCH and the number of vehicles expected to tune to a given SCHI. In the event of a high priority message during the SCHI, the invoking vehicle schedules the transmission of the emergency message across all the SCH's based on a schedule determined by the SCH which has the smallest fraction of the delay divided by the number of vehicles in the SCH. This implies that in WSD, the emergency message invoking vehicle performs the channel coordination function.

In this paper, the information collection routine during the CCHI based on the service advertisements is the same as that of the WSD except that each vehicle only collects the separation distance information between the vehicles in its communication range and the expected SCH they expect to tune to during the SCHI. We consider vehicles expecting to tune to similar specific SCHs as belonging to the same SCH cluster and for each SCH cluster a coordinator for each of the other SCH clusters is selected. The selection is based on the least average separation distance (LAD). This description of our scheme was first introduced in our paper [12]. Therefore, we extend the concept by; 1) detailing the proposed scheme, 2) performing an extensive literature survey of multi-channel MAC schemes in VANETs, 3) proposing a Markov chain for the back-off procedure during the SCHI, 4) a mathematical analysis of end-to-end delay which incorporates a proposed model for the optimal slot length when CMD operates during the CCHI, 5) and additional end-to-end delay performance tests in single hop blind flooding scenarios. The results of the study show that the proposed scheme has a lower end-to-end delay in both non rebroadcast scenarios and single hop flooding scenarios when compared to the WSD approach [6]. The original contributions of this article are summarized as follows:

- A multi-channel coordinator selection approach based on the LAD to vehicles tuned to other SCHs with the purpose of forwarding emergency messages with minimum end-to-end delay.
- A Markov chain for the back-off procedure during contention for transmission of safety messages in the SCHI.
- A model to determine the optimal slot length in which the proposed CMD operates during the CCHI.
- A queueing delay model that depends on the number of vehicles with in the carrier sensing range to determine the queue length.
- A mathematical analysis of the message dissemination end-to-end delay for the proposed CMD scheme and WSD.
- A simulation analysis of end-to-end delay while comparing the proposed CMD scheme, WSD and the legacy IEEE 1609.4.

The remainder of this paper is organized as follows. Section 2 discusses the related works. Section 3 describes the proposed CMD system model. Section 4 describes the numerical analysis. Section 5 describes shows the simulation setup and performance analysis. Finally, the conclusion is given in Section 6.

## 2. Related Work

Various state-of-the-art approaches designed for multi-channel VANET scenarios are discussed in this section. The review covers adaptive interval approaches and coordination based approaches used in multi-channel VANETs.

Pal *et al.* [3] proposed to eliminate the fixed CCHI and SCHI intervals by introducing a triggered multi-channel medium access control (MAC) scheme where the CCHI is triggered each time there exists an emergency message with the objective of minimizing the end-to-end-delay. Similarly, Chantaraskul *et al.* [13] and Wang *et al.* [5] also proposed approaches to dynamically adjust the CCHI based on the channel congestion condition. This approach offers a high trade-off against infotainment content delivery in environments where both safety and content delivery is highly is required.

Almohammedi *et al.* [4] proposed an adaptive multi-channel assignment and coordination (AMAC) scheme in VANETs which exploits channel access scheduling and channel switching. The channel access scheduling is done by the RSU based on the traffic conditions to guarantee that all safety messages are disseminated during the CCHI and also achieve higher throughput of the infotainment content. The AMAC scheme also uses a peer-to-peer (PNP) negotiation mechanism between service providers and users for the SCH reservations to adaptively transmit safety messages based on the CCH conditions and the traffic safety state. The PNP negotiation process results into; 1) transmission of safety messages over the CCH if the traffic condition is light; 2) transmission over the SCH if the traffic condition is heavy to avoid extended end-to-end delays of safety message delivery. Transmission over the SCH involves negotiating for a time slot during the SCHI. Generally, the PNP negotiation process is an additional process in the synchronization interval (SI) and naturally extends end-to-end delays. Additionally, AMAC uses different adaptive contention windows for safety message and service message transmission in order to minimize on packet collision in the multi-channel environment.

Similarly, Wang *et al*. [5] proposed a variable CCHI (VCI) multi-channel MAC which dynamically adjusts the length ratio between the CCH and the SCH mainly for the transmission of safety messages. In the VCI approach, when wireless service advertisements (WSAs) are transmitted during the CCHI, interested nodes request the service provider to reserve a specified content transmission time interval in the SCHI within which they shall receive content. This reservation approach is quite similar to the PNP time slot negotiated for in [4]. The only difference is that in [5], the time slot is used for transmitting infotainment content while in [4] the time slot is used for transmitting safety messages.

The hidden node problem in multi-channel VANETS can be minimized using the request to send (RTS) / clear to send (CTS) / data / acknowledgement (ACK) handshake. However, this causes the exposed node problem that hinders concurrent transmissions especially in dynamic environments like VANETs. In particular, SCH selection in multi-channel VANETS can result into an exposed node problem hence hindering concurrent transmissions. Lee *et al*. [8] proposed a scheme based on piggybacking of selected SCHs in the safety message in multi-channel VANETs to minimize the exposed node problem. In this case the piggybacked message acts as a coordination agent so that the exposed vehicles do not select a common SCH.

Yao *et al*. [9] proposed a flexible multi-channel MAC (FM-MAC) protocol which allows safety messages to be broadcasted on the service channel and non-safety messages to be transmitted on the control channel in a flexible way. The SCHI and CCHI are not adjusted dynamically but instead both are utilized for transmitting safety and non-safety messages. In FM-MAC, finding the optimal bandwidth resource allocation was key in determining the flexibility of using both the SCHI and CCHI. The RSU in [9] performs the major coordination function by; 1) setting up a coordination period for the RSU to broadcast frames to all vehicles in range informing them of a contention period to transmit safety messages 2) safety message broadcasts and SCH service reservation requests are made by vehicles 3) The RSU as well broadcasts a scheduling period to all vehicles in its range informing them of the schedule assignments and schedule orders. 4) and finally all non-safety messages are exchanged based on the SCH schedules and assignments which were broadcasted by the RSU. Zhao *et al*. [10] proposed the demand-aware MAC (DA-MAC) protocol which follows quite a similar criteria like in [9] though it DA-MAC does not consider the coordination frames broadcast by the RSU in FM-MAC.

The multi-channel coordination schemes in [9-11] seem attractive, but mainly depend on the RSU. It has been reported that RSUs may sometimes face unavailable grid power connection challenges [14] hence may require being battery powered. The major issue is ensuring that they are power charged. This limitation is the reason for the advocacy of vehicle-to-vehicle (V2V) target multi-channel coordination schemes.

The WSD algorithm proposed by Ghandour *et al*. [6] targets transmitting event driven high priority messages to all service channels with a minimized delay to its neighbours. During the CCHI, WSD operates at each node by gathering information about its neighbours through hello messages thereby forming a database comprising of the available service channels and available vehicles. In case there exists an emergency message event trigger during the SCHI, the SCH with the least average ratio of the channel average delay and the number of nodes is first tuned to by the source vehicle of the emergency message event trigger for message dissemination. SCH switching continues in the order of the least ratio until all SCHs are exhausted. The major point of interest in the WSD protocol is to disseminate information to its neighbours with minimum delay. Due to the multiple switches to different SCHs by the nearest vehicle which acts as a coordinator, WSD logically poses a large total dissemination delay in order to transmit to all the other service channels. The WSD design is based on the argument that nearer vehicles are a greater point of interest for safety.

The scheduling algorithm for high priority message dissemination (SAEMD) proposed by Joo *et al*. [15] operates by selecting and switching to a SCH belonging to the nearest vehicle. Similar to WSD in [6], SAEMD uses a data collection routine in the CCHI and uses the separation distance data for deciding on the nearest vehicles hence the next SCH to be tuned to for message transmission. Summarily, WSD [6] and SAEMD [15] were designed to work in multi-channel WAVE conditions. However, both WSD and SAEMD provide a minimum end-to-end delay benefit in the SCH which the nearest neighbouring vehicles tunes to first. In the case where most SCHs have vehicles tuned to them, the overall total dissemination delay is expected to be larger due to the need to do multiple switching to the different SCHs. Based on WSD and SAEMD, the total end-to-end delay for emergency message dissemination in multi-channel WAVE conditions needs to be improved. In our previous work [12], we presented a cooperative multi-coordinator scheme (CMD) for multi-channel communication in VANETs to take care of the large total dissemination delay in multiple service channels. The proposed CMD addresses multi-channel communications in VANETs and uses acquired knowledge from the CCHI.

Like in Dang *et al*. [16], the proposed CMD advocates for the utilization of the SCH in case an emergency message is invoked towards the time the SCHI takes over in the SI. Utilization of the both the CCHI and SCHI increases the reliability of safety message broadcasting. In the proposed CMD approach, each vehicle maintains a single radio, and the channel coordinator selection approach is distance based. CMD also makes use of multiple coordinators for each SCH cluster based on the available advertised SCHs here-after referred to as $Y$. Table 1 shows the comparisons of different state-of-the-art multi-channel access schemes used in VANETs.

**TABLE 1.** Comparison of existing multi-channel VANET schemes.

| Scheme | Utilizes RSU for coordination? | Nodes hosting the coordination function | Switching times per coordinator |
|---|---|---|---|
| Pal et al. [3] | No | 1 | $Y$-1 |
| Chantaraskul *et al.* [13] | No | 1 | $Y$-1 |
| VCI: [5] | No | 1 | $Y$-1 |
| AMAC: [4] | Yes | 1 | $Y$-1 |
| Lee et al. [8] | No | 1 | $Y$-1 |
| FM-MAC: [9] | Yes | 1 | $Y$-1 |
| DA-MAC: [10] | Yes | 1 | $Y$-1 |
| Li et al. [11] | Yes | 1 | $Y$-1 |
| WSD: [6] | No | 1 | $Y$-1 |
| SAEMD: [15] | No | 1 | $Y$-1 |
| Proposed CMD | No | $Y$-1 | 1 |

Switching times: refers to the number of times a coordinator node must switch to different SCHs to transmit a single emergency message until all SCHs receive the message.

## 3. Cooperative Multi-channel Emergency Message Dissemination Protocol (CMD)

CMD operates in vehicular multi-channel communications with the goal of achieving a low end-to-end delay in the dissemination of messages throughout the entire set of vehicles tuned to different SCHs without changing much on the IEEE 1609.4 standard. Like some of the presented multi-channel approaches in [6] and [15], the CMD protocol follows the channel coordination principle where the coordinator vehicles are selected using the distance to vehicles tuned to other SCHs. A channel coordinator selection algorithm is presented later in this section.

Figure 1(a) shows the standard IEEE 1609.4 channel access and Figure 1(b) shows the synchronization interval (SI) utilization based on the proposed CMD that can be described in the following steps:

1) At the start of the CCHI and after the guard interval, 26 ms are used for broadcasting basic safety messages (BSMs) and advertising available services by service provider nodes. The BSM's broadcasted at this stage includes the vehicle location information and the SCH that a node will use to in order to receive non-safety data.
2) In the next 5 ms, using the location information received and piggybacked SCHs from the other nodes, each node calculates the average distance it has from nodes which intend use each of the different SCHs, respectively.
3) The calculated average distance to vehicles intending to tune to each SCH is appended to the BSM and broadcasted by each vehicle. In the last 20 ms of the CCHI, the vehicles then broadcast their BSMs. On receipt of each BSM, each vehicle compares its own average separation distances with that in the received BSM if the SCH in the BSM is the same. A node autonomously qualifies itself to be the best fit coordinator if it has the LAD compared to all the other nodes intending to use the same SCH.
4) During the SCHI, in the event of an emergency event message transmission, the best-fit vehicles with the LAD to other SCHs forward the emergency message to the vehicles that tuned to the other SCH by switching to the target SCH.

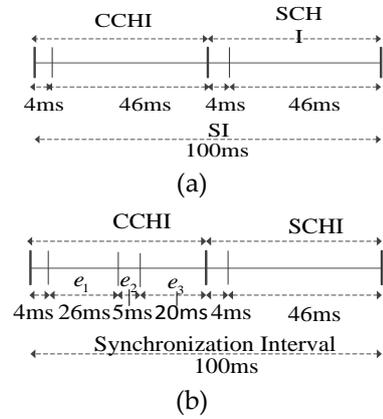

**FIGURE 1**: (a) Standard channel access in IEEE 1609.4 (b) The SI utilization based on CMD

### 3.1. Channel Coordinator Selection

In this subsection, the CMD protocol is described in detail and illustrated by Figure 2. Figure 2 shows three channels SCH1, SCH2 and SCH3 which were advertised during the CCHI and logically clustered to represent the vehicles tuned to the different SCHs during the SCHI. The vehicles selected the advertised SCHs in the CCHI in a random manner. Each vehicle while in the CCHI received and selected an SCH from the WAVE service advertisements (WSAs) and also received and transmitted location information together with their selected SCH. With the received location information and SCH at every instance, each receiving vehicle computes the separation distances in relation to each SCH with the objective of finding the least separation distance to vehicles expecting to tune to a specific SCH.

**TABLE 2.** Notations used in channel coordinator selection

| Acronym | Description |
|---|---|
| $k$ | An advertised SCH which a vehicle intends to switch to during the SCHI |
| $z$ | Any other advertised SCH apart from the one which a given vehicle intends to switch to during the SCHI |
| $c$ | An SCH coordinator vehicle |
| $c_{k\_z}$ | The channel coordinator for forwarding messages from SCH$k$ to SCH$z$ |
| $Y$ | The number of advertised SCHs to provide non-safety services |
| $m$ | The number of vehicles expecting to switch to SCH$z$ |
| $d_i$ | The V2V separation distance. $i = 1,...,m$ |
| $d_{i\_z}$ | The average $d_i$ for a given vehicle considering the vehicles expecting to switch to SCH$z$ |
| $D_{c\_z}$ | The LAD for the coordinator vehicle in SCH$k$ to SCH$z$ |
| $f_{SCHz}$ | The coordination fitness value for a given vehicle considering the $d_{i\_z}$ to SCH$z$ |

Considering each SCH as a cluster, the channel coordinator vehicles in each cluster are such that for all vehicles in a given cluster, they have LAD of the connectivity to nodes in another SCH compared to the other vehicles it will share with the same SCH. If $Y$ SCHs were advertised, then there should exist $Y-1$ SCH coordinators in each cluster. Table 2 describes the notations used in formulating the channel coordinator selection approach. The channel coordinator selection model can be formulated as

$$\exists C_{k\_z} \in \text{SCH}k \text{ s.t. } D_{c\_z} \le d_{i\_z} \ \forall d_{i\_z}, \ i=1,2,.....m, \quad (1)$$
$$k = 1,2,3....,6.$$

where

$$d_{i\_z} = \frac{d_1 + d_2 + ... + d_m}{m} \text{ for } z = 1,2,3...,6 \quad (2)$$

Each vehicle keeps the broadcasted SCH$z$ and their associated $d_{i\_z}$ in its coordination fitness information base (CFIB) as seen in Table 3. After the $d_{i\_z}$ calculation stage by each receiving vehicle, each vehicle again broadcasts its local $d_{i\_z}$ and is received through the periodic broadcast BSM. Each incoming $d_{i\_z}$'s are compared with the local $d_{i\_z}$'s as long as the SCH$z$ is the same. The comparison is such that when $d_{i\_z}$ is the least among the incoming $d_{i\_z}$'s for the common SCH$z$, then the coordination fitness (CF) is 1. Implying that the vehicle $i$ has the least average distance to SCH$z$ and hence is the service coordinator of its SCH to SCH$z$. Generally, the value of CF is determined based on the order of greatness of $d_{i\_z}$. That is, the least $d_{i\_z}$ has CF

**Algorithm 1:** CHANNEL COORDINATOR SELECTION ALGORITHM

1. **while** in CCHI vehicles receive WSA's and broadcast their location information
2. Select an SCH to be tuned to
3. Append selected SCH and location information to all BSM's and broadcast
4. **while** periodic safety messages are received
4.a    **for** each vehicle
4.b       **for** each SCH advertised
4.c          Compute $d_{i\_z}$
4.d          Append $d_{i\_z}$ to the BSM and then broadcast
         e**nd for**
4.e       **if** ( BSM is received ) **then**
            **for** each similar SCH$z$
4.f               **if**(all the $d_{i\_z}$ values are greater than the local average $d_{i\_z}$ ) **then**
4.g                Vehicle is the channel coordinator $C_{k\_z}$
4.h          **else**
4.i                Vehicle is just a member of its selected SCH$k$ cluster.
4.j          **end if**
            **end for**
4.k       **end if**
4.l    **end for**
4.m  **end while**
5. **end while**

**TABLE 3.** Coordination fitness information base.

| Gossiped SCH$z$ | Average $d_{i\_z}$ | $f_{SCHz}$ |
|---|---|---|
| SCH 1 | $d_{i\_1}$ | $\ge 1$ |
| SCH 2 | $d_{i\_2}$ | $\ge 1$ |
| SCH 3 | $d_{i\_3}$ | $\ge 1$ |
| SCH 4 | $d_{i\_4}$ | $\ge 1$ |
| SCH 5 | $d_{i\_5}$ | $\ge 1$ |
| SCH 6 | $d_{i\_6}$ | $\ge 1$ |

equals to 1 and the greatest $d_{i\_z}$ has CF equals to $m$. For clarity, the CF range is $i = 1, 2, 3, ..., m$. The least $d_{i\_z}$ which represents the coordinators average distance is then represented as $D_{c\_z}$ for purposes of clarity as seen in (1).

Again, as seen in Table 3, the CF value in **SCH**$k$ is represented as $f_{SCHz}$. The general representation in Figure 2 shows the vehicle coordinators $C_{k\_1}, C_{k\_2}, C_{k\_3}$ which have the least CF values to the advertised SCHs. It should however be noted that although $C_{2\_3}$ and $C_{3\_2}$ are represented a SCH coordinators in Figure 2, only $C_{1\_3}$ and $C_{1\_2}$ are functionally operational as channel coordinators because the emergency message is triggered in SH1. Algorithm 1 elaborates on the CMD channel coordinator selection procedure.

### 3.2. Challenges in the Proposed CMD

In the CCHI, while transmitting BSM's containing the average separation distance to other vehicles, conditions such as the hidden node problem and shadowing may hinder the BSM delivery to some vehicles. In such a case, more than one vehicle may assume the position of the channel coordinator to a given SCH cluster. During the SCHI, it is also possible that a channel coordinator vehicle may not receive an emergency message from the affected source vehicle due to the hidden node problem.

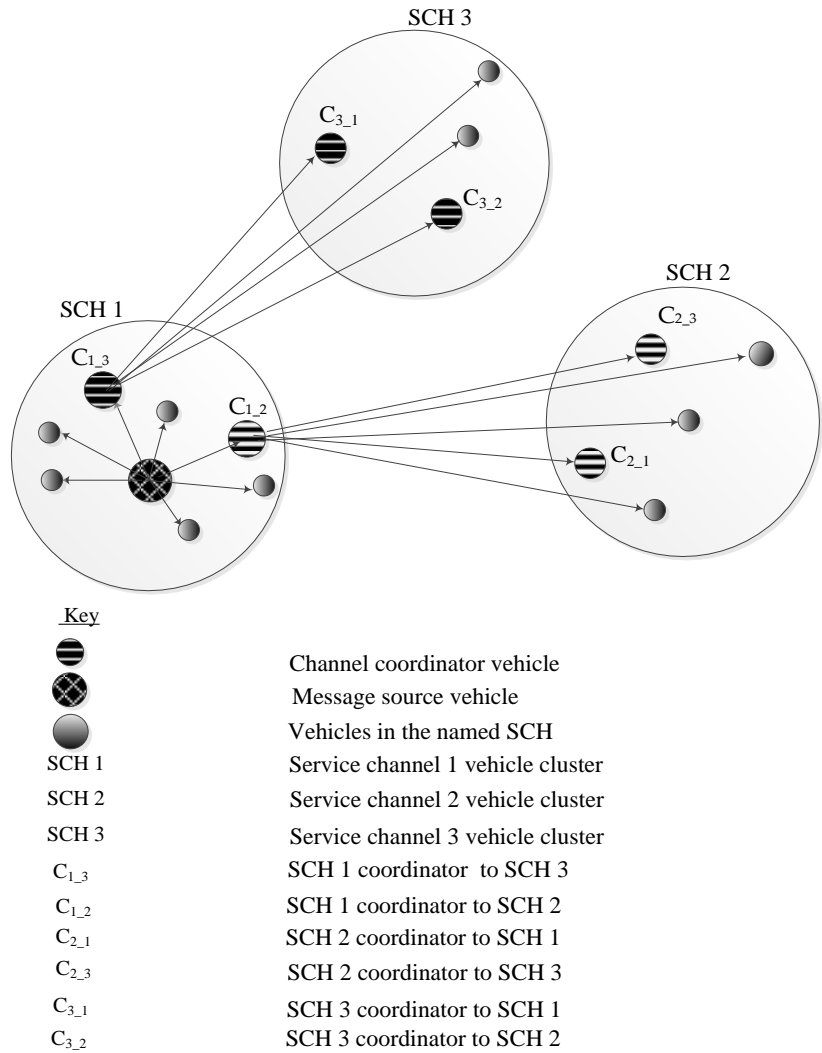

Key:
- Channel coordinator vehicle
- Message source vehicle
- Vehicles in the named SCH
- SCH 1 — Service channel 1 vehicle cluster
- SCH 2 — Service channel 2 vehicle cluster
- SCH 3 — Service channel 3 vehicle cluster
- $C_{1\_3}$ — SCH 1 coordinator to SCH 3
- $C_{1\_2}$ — SCH 1 coordinator to SCH 2
- $C_{2\_1}$ — SCH 2 coordinator to SCH 1
- $C_{2\_3}$ — SCH 2 coordinator to SCH 3
- $C_{3\_1}$ — SCH 3 coordinator to SCH 1
- $C_{3\_2}$ — SCH 3 coordinator to SCH 2

**FIGURE 2:** A logical view of the CMD structure with the emergency message generated from SCH1 and broadcast to its members then relayed by the channel coordinators to SCH3 and SCH2.

This is a prominent problem in single hop broadcast scenarios. To alleviate this reachability problem, the single hop blind flooding based approach of broadcasting was implemented and simulation results shown later in Subsection 5.4 to describe its impact on delay in each WAVE channel. In single hop blind flooding, when vehicles receive a message, they rebroadcast it only once. That is, the vehicles receiving the rebroadcasted message do not broadcast the retransmitted message. A comparison of the proposed CMD with WSD is also done for the single hop flooding scenario.

Again, by applying CMD, it is possible that only one in a given SCH may qualify to be the channel coordinator to all other SCHs by having the LAD to all advertised SCHs. This scenario exists when one node is isolated from its SCH members yet near to all the other SCH cluster members. Another issue about CMD is that when a cluster has less than $Y - 1$ members, then some members will act as coordinators for more than one SCH. These two mentioned scenarios would cause an increase in the total dissemination delay because such coordinators will have to switch between multiple channels.

### 3.3. Proposed Back-off Model for Emergency Message Transmission during the SCHI

Figure 3(a) represents the standard back-off process to be adopted in the CCHI and for non-safety data transmission in the SCHI. The Markov chain proposed and presented in Figure 3(b) operates in the SCHI showing the back-off process when an emergency message is invoked. Safety emergency messages are considered high priority messages during the SCHI therefore the model design is tailored to minimize their contention delay. In the standard back-off criteria, waiting state transitions are marked by uniformly reducing contention window sizes.

In the proposed criteria seen in Figure 3(b) the same phenomenon is followed but the size of the reducing contention window (RCW) is two times the size of the RCW compared to when transmitting WSA's, safety messages in the CCHI, and data services during the SCHI.

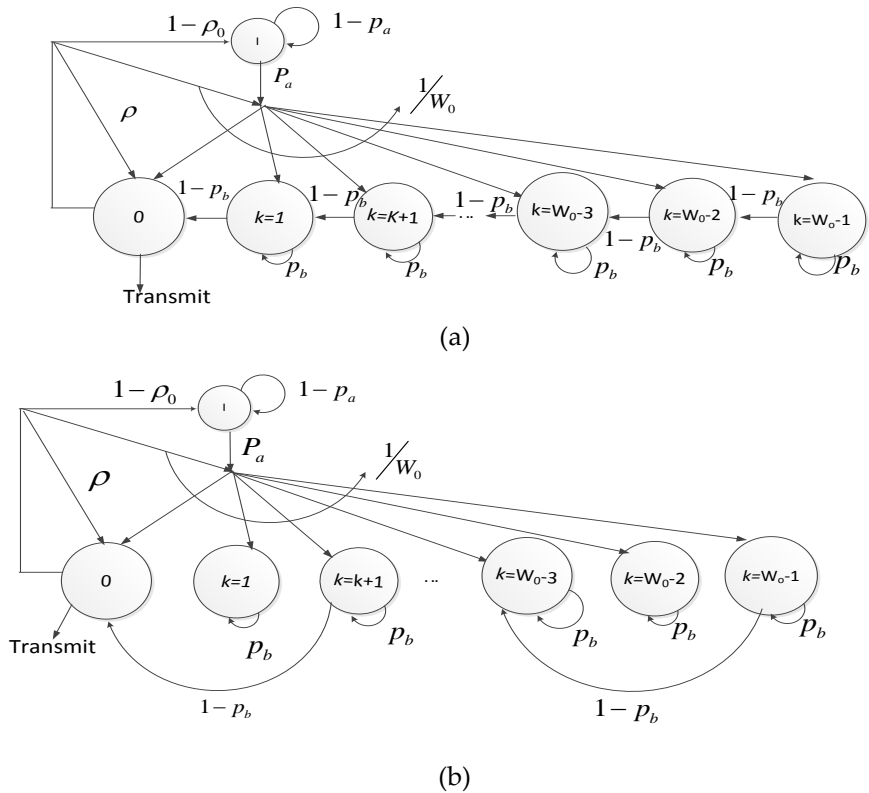

**FIGURE 3:** One-dimensional Markov chain model for a back-off instance. (a) Standard back-off process to be adopted in the CCHI and for non-safety data transmission in the SCHI (b) Proposed back-off process for emergency safety message transmission during the SCHI

**TABLE 4:** Notations

| Acronym | Description |
|---|---|
| $\lambda$ | The packet arrival rate. |
| $\mu$ | Average service rate of the queue in packets per second. |
| $\rho$ | The probability that at least one packet is in the queue = $\lambda/\mu$. |
| $p_b$ | Is the back-off blocking probability. |
| $p_a$ | The packet arrival probability $= 1 - e^{-\lambda\sigma}$. |
| $W_0$ | Contention window size for back-off. |
| $k$ | Current window size state as an effect of exponential back-off. |
| I | Idle state. |
| $\beta$ | Traffic density |
| $d_0$ | The reference distance used in calculating the received signal strength at a particular distance |
| $P_r(.)$ | The received signal strength at specified distance |
| $d_c$ | The critical distance that refers to the distance where the first Fresnel zone touches the ground and is also referred to as the Fresnel distance. |
| $\gamma_1$ | Path loss exponent |
| $\gamma_2$ | Path loss exponent |
| $h_T$ | Transmitter height |
| $h_R$ | Receiver height |
| $\psi$ | Electromagnetic wavelength fixed at 5.9 GHz |
| B | The number of vehicles in carrier sensing range |
| $X_{\sigma 1}$ | The zero mean, normally distributed random variables with standard deviation $\sigma 1$ |
| $X_{\sigma 2}$ | The zero mean, normally distributed random variables with standard deviation $\sigma 2$ |
| $L_{CS}$ | the carrier sensing range defined as the average distance for a node to detect the other nodes transmissions |
| $c_{th}$ | The carrier sensing threshold which indicates the receive sensitivity of the radio and is a constant and radio dependent. |

Let $s_i(t)$ and $b_i(t)$ represent the back-off stage and the back-off counter respectively at time $t$. Hence, the state of the Markov chain can be expressed as a two-tuple $\{s_i(t), b_i(t)\}$, and the back-off state of the high priority emergency messages can be simplified as a one-tuple $\{b_i(t)\}$ for $s_0(t) \equiv 0$. Table 4 defines all the probabilities shown in the Markov chains in Figures 3 (a) and 3(b). Each of the one-time transition probabilities in Figure 3(b) is described below:

- The idle state $\{I\} \rightarrow$ the back-off state $\{0\}$: Node transmits a packet if the channel is sensed as idle: $P\{0|I\} = p_a$.
- The idle state $\{I\} \rightarrow$ the state back-off $\{k\}$: This occurs if a new packet arrives in the queue: $P\{k|I\} = p_a/W_i$, $k \in (0, W_0 - 1)$.
- The back-off state $\{k\} \rightarrow$ the state back-off $\{k\}$: Occurs if the channel is sensed to be busy and in this case the back-off counter freezes: $P\{k|k\} = p_b$, $k \in (1, W_0 - 1)$.
- The back-off state $\{k+2\} \rightarrow$ the state back-off $\{k\}$: If the channel is sensed to be idle, the back-off counter decrements by two steps: $P\{k|k+2\} = 1 - p_b$, $k \in (0, W_0 - 2)$.
- The back-off state $\{0\} \rightarrow$ the idle state $\{I\}$: Node returns to idle state if it has no packet to send: $1 - \rho$.
- The back-off state $\{0\} \rightarrow$ the idle state $\{k\}$: Nodes starts back-off procedure if at least one packet is in the queue: $P\{k|0\} = \rho_0/W_0$, $k \in (0, W_0 - 1)$.

In summary, the one-step transition probabilities are as

follows

$$\begin{cases} P\{0 \mid I\} = p_a \\ P\{k \mid I\} = p_a / W_i, \ k \in (0, W_0 - 1) \\ P\{k \mid k\} = p_b, \ k \in (1, W_0 - 1) \\ P\{k \mid k + 2\} = 1 - p_b, \ k \in (0, W_0 - 2) \\ P\{I \mid 0\} = 1 - \rho_0 \\ P\{k \mid 0\} = \rho_0 / W_0, \ k \in (0, W_0 - 1) \end{cases} \quad (3)$$

The stationary distribution of the Markov chain is defined as

$$b_0 = \lim_{t \to \infty} P\{b(t) = k\}, \ k \in (0, W_0 - 1) \quad (4)$$

Given the one-step probabilities, the stationary probabilities can be expressed as

$$b_k = \frac{(W_0 - k)}{W_0 (1 - p_b)} b_0 \quad (5)$$

$$b_I = \frac{(1 - \rho)}{p_a} b_0 \quad (6)$$

The sum of the stationary probabilities for the states should be equal to one, therefore,

$$\frac{(W_0 - 1)}{W_0 (1 - p_b)} b_0 + \frac{(1 - \rho)}{p_a} b_0 = 1 \quad (7)$$

$$b_0 = \left[ \frac{(W_0 + 1)}{2(1 - p_b)} + \frac{1 - \rho}{p_a} \right] \quad (8)$$

Since transmission occurs when the back-off counter value $k = 0$, the transmission probability $\tau$ can be defined as:

$$\tau = b_0 = \left[ \frac{(W_0 + 1)}{2(1 - p_b)} + \frac{1 - \rho}{p_a} \right] \quad (9)$$

$\tau$ is very important as it is later used in the end-to-end delay analysis seen in the next section.

## 4. End-to-end Delay Analysis

The key performance indicator in this study is end-to-end delay. The goal of this section is to numerically derive the end-to-end delay while considering the mechanism of the proposed CMD scheme. Generally, the performance of the proposed CMD depends on the communication performance during the 26 ms of transmitting the location information and then the 20 ms of sharing the average separation distances to determine the SCH coordinators. The two decision time slots (26 ms and 20 ms) in this article from now-on-wards shall be referred to as $e_1$ and $e_3$ respectively as shown in Figure 1 (b).

Most importantly, all or most of the vehicles should transmit their information within $e_1$ and $e_3$ for the channel coordinator selection to be efficient. Therefore, one eminent optimization parameter in this problem is the length of $e_1$ and $e_3$ which we believe should depend on the length of an arbitrary time slot $T_{slot}$ exists during the interval $e_1$ and $e_3$. And since $T_{slot}$ is one parameter that determines the end-to-end delay of a transmission, we start by defining the end-to-end delay $\mathrm{E}[d]$ model as follows

$$\mathrm{E}[d] = \mathrm{E}[q] + \mathrm{E}[c] + \mathrm{E}[t] \quad (10)$$

Where $\mathrm{E}[q]$, $\mathrm{E}[c]$ and $\mathrm{E}[t]$ represent the average queueing delay, average contention delay and average transmission delay, respectively.

### 4.1. Contention Delay Model

The average contention $\mathrm{E}[c]$ is defined as

$$\mathrm{E}[c] = \mathrm{E}[CW] = ((CW_{\min} - 1)/2) T_{slot} \quad (11)$$

where $\mathrm{E}[CW]$ is the average contention window size. The size of $T_{slot}$ is relevant for the derivation of the optimal period for $e_1$ and $e_3$ for the proposed CMD. Finding $T_{slot}$ requires that; 1) we define the stationary probability that a node transmits a BSM in the arbitrary time slot $T_{slot}$ 2) the time it takes to yield a successful transmission $T_{success}$, collision time $T_{coll}$ and the idle time $\sigma$.

By using the transmission probability $\tau$, the following probabilities can be found

$$\begin{aligned} p_{idle} &= (1 - \tau)^N \\ p_{busy} &= 1 - p_{idle} \\ p_{success} &= N\tau (1 - \tau)^{N-1} \\ p_{coll} &= 1 - p_{idle} - p_{success} \end{aligned} \quad (12)$$

where $p_{idle}$ is the probability that a channel is in an idle state and not being utilized, $p_{busy}$ is the probability that a transmission is occupying the channel, $p_{success}$ is the probability of having a successful transmission and $p_{coll}$ is the probability of having a collision in the channel.

The transmission time slot duration $T_{slot}$ is defined as:

$$T_{slot} = (1 - p_{busy})\sigma + T_{success} \cdot p_{success} + T_{coll} \cdot p_{coll} \quad (13)$$

where σ is the duration of an empty slot. $T_{success}$ is the time required for a successful transmission, and $T_{coll}$ is the average time of a collision event.

$$T_{success} = \text{DIFS} + \sigma + \text{E}[t] \quad (14)$$

$$T_{coll} = \text{EIFS} + \sigma + \text{E}[t] \quad (15)$$

The average transmission delay can be expressed as $\text{E}[t] = S/R$, with $S$ representing the message size and $R$ representing the data rate, respectively. DIFS and EIFS are the distributed coordination function inter frame space time and extended inter frame space time respectively.

**4.1 Optimal Slot Period Allocation Model**

At this stage, since $T_{slot}$ has been mathematically defined by equation (13), the task is now to define the optimal period of that each of $e_1$ and $e_3$ slot shall take. In other words, we need to find how many $T_{slot}$'s should exist in either the 1st or 2nd time slot to enable sufficient coordination selection functionality

The objective to achieve during $e_1$ and $e_3$ is to have most or all of the vehicles to transmit their location, desired SCH and LAD information. In this article, we consider that $e_1$ and $e_3$ period should just be long enough to allow all the vehicles denoted by $B$ within the carrier sensing range to transmit their information. The duration $V$ representing either $e_1$ or $e_3$ can therefore be defined as:

$$V = B \times T_{slot} \quad (16)$$

In this article, we define the number of vehicles $B$ in carrier sensing range based on [9] as:

$$B = 2\beta L_{cs} \quad (17)$$

$L_{cs}$ is given by:

$$L_{cs} = \begin{cases} E[d_0 10^{\frac{p_r(d_0)-c_{th}+X_{\sigma 1}}{10\gamma_1}}], & d_0 \leq L_{cs} \leq d_c \\ E[d_c 10^{\frac{p_r(d_0)-10\gamma_1 \log_{10}(d_c/d_0)-c_{th}+X_{\sigma 2}}{10\gamma_1}}], & L_{cs} > d_c \end{cases} \quad (18)$$

$d_c$ can be calculated as $d_c = \frac{4h_T h_R}{\psi}$.

**4.2. Queueing delay model**

In this paper, the queueing delay $\text{E}[q]$ is formulated considering that a VANET communication system is best modeled as an M/M/1/B queueing system [17]. In this case, the arrivals are considered to be distributed exponentially through a Poisson process, the service times are exponentially distributed and independent of each other, a single communication channel acting as a server and has a finite queue length $B$. Where we define $B$ in this article as the number of vehicles within the carrier sensing range. Based on equation (17), $B$ can be calculated. The expected queue length can therefore be calculated as

$$\text{E}[b] = \frac{\rho}{1-\rho^{B+1}} \cdot \left(\frac{1-\rho^B}{1-\rho} - B\rho^B\right) \quad (19)$$

Using Little's law, the queueing delay can be represented as

$$Q_d = E[b]/\lambda(1-P_B) \quad (20)$$

Where $P_B$ is the probability that the queue is full and $\lambda(1-P_B)$ represents the effective arrival rate which the packets are put into the queue. When $\rho = (\lambda/\mu) \neq 1$, the queueing delay is defined as

$$\text{E}[q] = Q_d = \frac{E[b]}{\lambda\left(1 - \frac{1-\rho}{1-\rho^{B+1}} \cdot \rho^B\right)} = \frac{1}{\mu-\lambda} - \frac{1}{\mu} \cdot \frac{B\rho^B}{1-\rho^B} \quad (21)$$

when
$\rho = 1$, $Q_d = E[b]/\lambda\{1-[1/(B+1)]\} = (B+1)/2\lambda = (B+1)/2\mu$

At this stage, all the parameters for numerically finding $\text{E}[d]$ using (10) can be computed.

**5. Simulation**

**5.1 Mobility Model and Network Simulator**

The Manhattan model is used to emulate the movement pattern of vehicle nodes on streets defined by a map. The map is composed of a number of horizontal and vertical streets. Each street has one lane. The mobile vehicle node moves along the horizontal and vertical grids on the map. At an intersection of a horizontal and vertical streets, the mobile node can turn left, right, or goes straight. This choice is probabilistic. The vehicle turn probability is set to 0.5. We consider a two-dimensional 1,500 m by 1,500 m fully connected road network in a Manhattan grid with vehicles moving at a mean speed of 40 km/h. The grid offers a total of 6 km for vehicular motion for the single-lane scenario. Our mobility trace for the vehicles is generated using BonnMotion-2.1.3.

To analyze the performance of CMD, we simulated its system dynamics with the NS-3 simulator, version ns-3-dev. Table 5 summarizes the general simulation parameters and Table 6 defines the simulation performance metrics.

**5.2 End-to-end Delay**

In a typical VANET scenario, not all vehicles may demand for the advertised infotainment services. This means that not all SCH's will be utilized during the SCHI. In Figure 4, the total end-to-end dissemination delay is shown for WSD, IEEE 1609.4 and the proposed CMD. Only 5 SCHs were advertised during the CCHI.

Observations show that the proposed CMD maintains lower total end-to-end delays compared to WSD and the legacy IEEE 1609.4 when more than two SCHs are utilized during the SCHI. This observation is true for both the analytical and simulation results. In the legacy IEEE 1609.4, a vehicle with an emergency message during the SCHI must wait for the CCHI in order to transmit an emergency message. This is the major cause for the much end-to-end delay exhibited by the legacy IEEE 1609.4 system. The better performance realized by the CMD is the effect of using multiple coordinators whereby each coordinator switches to a specific SCH in order to relay a BSM during the SCHI. In WSD, only one channel coordinator is used, hence the need for multiple channel switching in order to relay the BSM to all the SCH's. Therefore, there is an additional delay introduced by the multiple switching and the transmission delays.

The slight differences seen in the theoretical and simulation results are a result of the system dynamics used in generating the results both in theory and in the simulation. In WSD the theoretical results are generated based on the derivation of a single channel end-to-end delay $E[d]$. We then use the number of SCHs $Y$ as a factor to fix the multi-channel condition to find the total message dissemination end-to-end delay $T_d$ as follows

$$T_d = \begin{cases} E[d], & Y = 1 \\ Y E[d], & Y > 1 \end{cases} \quad (22)$$

In the proposed CMD, the theoretical $T_d$ is defined by

$$T_d = \begin{cases} E[d], & Y = 1 \\ 2 E[d], & Y > 1 \end{cases} \quad (23)$$

In the simulation, the frequency of each of the SCHs defined by the WAVE standard is different. This has an impact on the end-to-end delay results thus causing the slight differences observed between the theoretical and simulation results. It should be noted that the final $T_d$ represented in the results of Figures 4, 7 and 10 includes the switching delay where multiple channels are involved. Theoretically, the switching delay was arbitrarily fixed at 2 ms.

**5.3 PRR and PTR**

The proposed CMD first operates during the CCHI within the time durations, $e_1$, $e_2$ and $e_3$. During the time durations $e_1$ and $e_3$, it is important that all or most vehicles transmit and receive the BSM's in order to enable efficient channel coordinator selection. Therefore, Figure 5 is shown to provide an understanding of the PRR and the PTR during the time intervals $e_1$ and $e_3$.

**TABLE 5:** Simulation parameters.

| Description | Value |
|---|---|
| Message payload size $S$ | 200 bytes |
| Fading model | Nakagami |
| Packet interval | 100 ms |
| Data rate $R$ | 3 Mbps |
| Content window size- Min, max | 15, 256 |
| Slot time $\sigma$ | $16\,\mu s$ |
| Arbitrary inter-frame space number (AIFSN) | 2 |
| Short inter-frame space (SIFS) time | $32\,\mu s$ |
| Antenna height | 1.5 m |
| Frequency | 5.9 GHz |
| Transmitter and Receiver gain | 3 dB |
| Number of vehicles | 50 |
| Vehicle speed | 40 m/s |
| Vehicle mobility model | Manhattan-grid highway |

**TABLE 6:** Simulation performance metrics.

| Metric | Description |
| --- | --- |
| End-to-end delay | The safety message dissemination single-hop delay |
| Packet reception ratio (PRR) | The percentage of nodes that successfully receive a packet from a tagged node given that all the receivers are within the transmission range of the sender at the moment that the packet is sent out [18]. |
| Packet transmission ratio (PTR) | The percentage of nodes that successfully transmit a packet given the prevailing contention for channel access. |

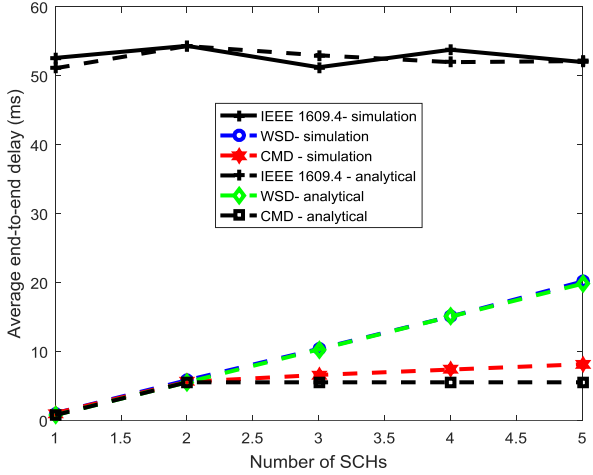

**FIGURE 4:** Analytical and simulation results of average end-to-end delay versus number of channels.

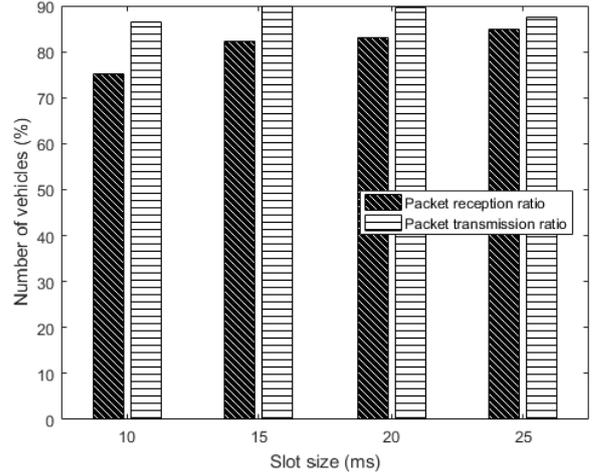

**FIGURE 5:** PRR and PTR simulation results for various sizes of $e_1$.

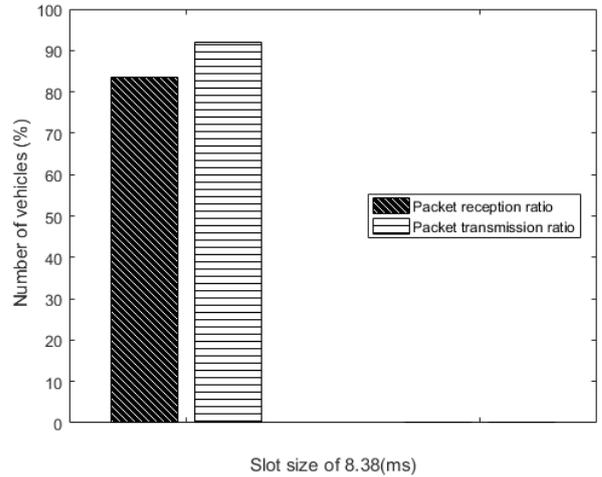

**FIGURE 6:** PRR and PTR simulation results based on the derived optimal $e_1$ interval.

It is observed in Figure 5 that as the slot duration of $e_1$ or $e_3$ increases, the PRR and PTR also increases. Generally, an increase in the slot duration gives room for more contending nodes to transmit as the available transmission time slots $\sigma$ would also increase.

Figure 6 represents the PRR and PTR realized when the proposed optimal $e_1$ model is used. The optimal length in time for $e_1$ is 8.38 ms given the simulation scenario and settings seen in Table 7. The parameter settings seen in Table 6 are based on realistic channel measurements which were attained in [19].

The key observation in Figure 5 and Figure 6 is that, $e_1$ values greater than 8.38 ms result into relatively the same PRR and PTR values with insignificant differences. This therefore means that lengthening $e_1$ or $e_3$ beyond 8.38 ms would simply be a waste in the CCHI.

**TABLE 7:** Parameter settings for optimal $e_1$ determination.

| Description | Value |
| --- | --- |
| $d_0$ | 10 m |
| $P_r(d_0)$ | -60 dB |
| $c_{th}$ | -85 dB |
| $X_{\sigma 1}$ | 5.6 dBm |
| $\gamma_1$ | 1.9 |
| $\beta$ | 25 vehicles/km |

Figure 7 represents the PRR attained against the total end-to-end delay achieved when transmitting a BSM over single and multiple SCH's. The result shows that the proposed CMD offers a greater PRR within a shorter end-to-end delay compared to the WSD and IEEE 1609.4 legacy system especially when considering total coverage of all SCH's with the BSM. The order of the SCH switching represented in Figure 7 for each approach depends on the channel switching dynamics of each.

At about 6 ms, CMD covered slightly over 50% of the vehicles and served 3 SCHs while WSD served lesser. The good performance exhibited by CMD is based on the multi-coordinator functionality in a scenario where multiple services are demanded and offered by different SCH's. It is important to note again that the IEEE 1609.4 would wait for the CCHI to transmit BSM's in case of an emergency during the SCHI. It is for this reason that the end-to-end delay for the legacy system is not better than CMD and WSD.

## 5.4 Improving Reachability for Reliability by Single Hop Blind Flooding

In order to provide insights on how to alleviate the hidden node problem which can be a hindrance to the effectiveness of the proposed approach during the channel coordinator selection process, we have implemented the single hop bind flooding approach well knowing that blind flooding approaches introduce the broadcast storm problem [20] which may affect the end-to-end delay.

The purpose of experimenting the single-hop blind flooding (SHBF) is to provide an understanding that even though using SHBF introduces further end-to-end delays, it can be used as a factor in further determining the optimal size of $e_1$ and $e_3$ with the benefit of having a higher reachability during $e_1$ and $e_3$.

However, in this study, we have not divulged into further formulating another model for determining the optimal size of $e_1$ and $e_2$ based on the SHBF end-to-end delay results. We only present SHBF based results.

Figure 8 shows the cumulative distribution function of the reachability in both flooding and no flooding conditions in the control CCHI given a period of 8 ms. The results captured in Figure 8 are for the first SI in our simulation experiment particularly to understand the influence of the number of vehicles in the simulation playground especially given the fact that the vehicle node generation in the simulation is based on a Poisson process.

Four sections of reachability for analysis can be observed in Figure 8. These are between 0 and 10%, between 10% and 38%, between 38% and 68%, and > 68%.

The reachability range between 0% and 10% is realized during the starting period of the SI when few vehicle nodes have been ushered into the simulation environment based on a Poisson process. It can be observed that the no-flooding scenario offers a better reachability compared to the SHBF scenario. This is because at the start there are few vehicles which are all able to be reached and therefore, introducing the SHBF simply causes unnecessary contention.

As the number of vehicles increases in the simulation environment, the sparsity of the vehicles is larger given the vehicle mobility. This sparsity leads to reduced reachability. This can be observed between 10% and 38% where the SHBF scenario offers a better reachability compared to the no-flooding scenario.

The number of vehicles in the simulation environment increases to a point where by there is a level of stability in the reachability which can be observed between 38% and 68%. This stability scenario is true for both the SHBF and the non flooding scenario. This means that SHBF has no effect in the CMD process in dense vehicular scenarios.

After 68% reachability is achieved, using the SHBF scenario does not offer better reachability results because of the broadcast storm. At this moment, all vehicles are in the playground of the simulation environment.

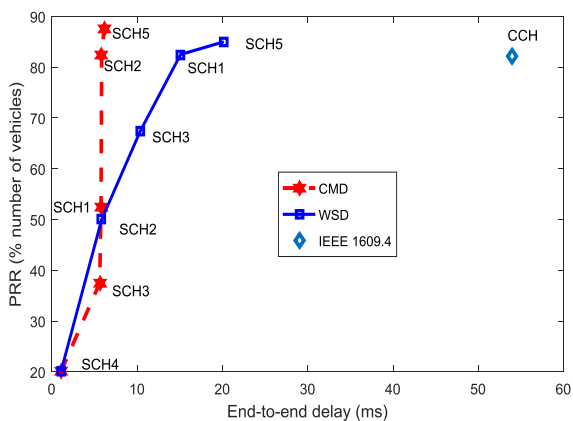

**FIGURE 7:** PRR versus end-to-end delay: Understanding the BSM proliferation rate across various channels.

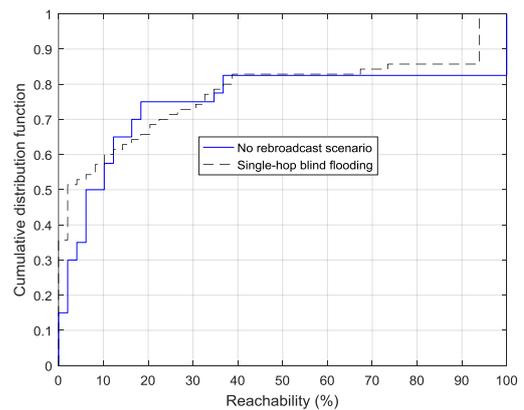

**FIGURE 8:** Cumulative distribution function of the percentage number of vehicles receiving message transmission during the CCHI

We can generally affirm from the observations that the SHBF is indeed suitable to improve on reachability in sparsely dense vehicular scenarios as seen in the region between 10% and 38%. Therefore, the SHBF is useful in the CMD process in sparsely dense vehicular scenarios.

To investigate the effect of flooding on delay, the single hop blind flooding was implemented in five WAVE SCH's with the objective that during the SCHI, there should be a higher guarantee of emergency message delivery to the channel coordinator once invoked by any vehicle.

By observing Figure 9, it is clear that the single hop blind forwarding introduces a further delay in the message dissemination time compared to when no blind flooding is applied.

Observations in Figure 10 also indicate that as a result of single hop blind flooding, the average total dissemination end-to-end delay over multiple channels will also increase compared to what was earlier realized in Figure 5 when no flooding was applied. However, it is worth noting that in scenarios of no flooding and single hop blind flooding, CMD still exhibits a delay lesser than WSD which is desirable for our design goal.

The negative impact of single hop blind flooding observed in Figures 9 and 10 imply that a good minimum delay flooding mechanism once utilized would further improve the performance of our proposed CMD protocol in the process of disseminating BSM's.

## 6 Conclusion

In this paper, we proposed a cooperative multi-channel message dissemination scheme called CMD for safety message dissemination in the IEEE 1609.4 standard with the goal of improving on the reliability of safety messaging in multi-channel scenarios. In order to achieve this; a cooperative SCH coordinator selection approach was developed. The SCH coordinator selection is based on the vehicle which has the LAD to vehicles that expect to tune to other SCH's and operates during the CCHI.

In order to improve on the efficiency of the channel coordinator selection process during the CCHI, a model to determine the optimal slot duration was developed. A channel contention back-off Markov model was developed to operate during the SCHI in order to improve on the transmission of high priority safety messages in the event that they are invoked. Additionally, a queueing delay model that depends on the number of vehicles within the carrier sensing range was proposed and developed to determine the queue length.

Through mathematical and simulation analysis, the proposed CMD achieves lower end-to-end delay and PRR compared to the legacy IEEE 1609.4 system and WSD, which is one of the state-of-the-art multi-channel schemes for WAVE.

## Data Availability

The data used to support the findings of this study are available from the corresponding author upon request.

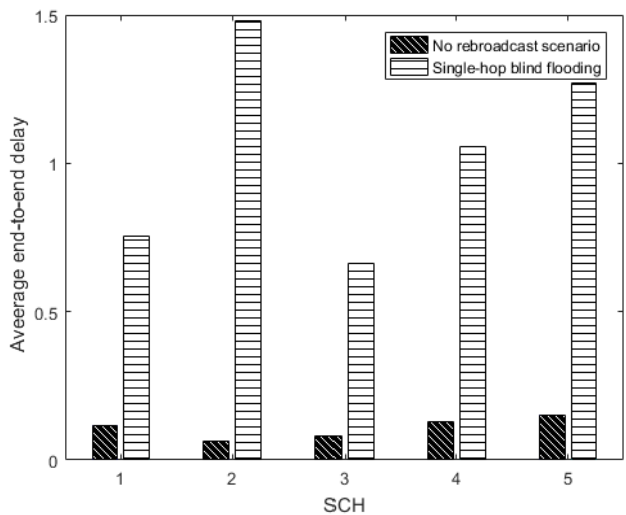

**FIGURE 9:** Average dissemination delay in each channel while comparing the blind flooding scenario with the non rebroadcast scenario at each SCH.

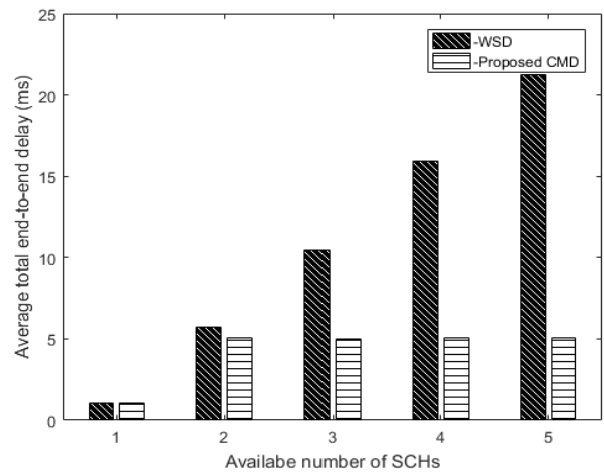

**FIGURE 10:** Average total dissemination delay in the single hop flooding scenario given varying numbers of available SCH's.

## Conflicts of Interest

The authors declare that there are no conflicts of interest regarding the publication of this paper.

## Acknowledgements

This work was partially supported by Basic Science Research Program through the National Research Foundation of Korea (NRF) funded by the Ministry of Education (NRF-2016R1D1A3B03934420) and the Korea Institute for Advancement of Technology (KIAT) grant funded by the Korean government (MOTIE). (No.P0000535, Multichannel telecommunications control unit and associated software).